\input{aipcheck}

\documentclass[final]{aipproc}
\layoutstyle{6x9}
\begin{document}

\title{Leptogenesis, Dark Energy, Dark Matter and the neutrinos}
\author{Utpal Sarkar\footnote{E-mail:utpal@prl.res.in}}{
  address={Physical Research Laboratory, Ahmedabad - 380 009, INDIA}
}
\begin{abstract}

In this review we discuss how the models of neutrino masses can
accommodate solutions to the problem of matter-antimatter 
asymmetry in the universe, dark energy or cosmological constant
problem and dark matter candidates. The matter-antimatter asymmetry
is explained by leptogenesis, originating from the lepton number 
violation associated with the neutrino masses. The dark
energy problem is correlated with a mass varying neutrinos,
which could originate from a pseudo-Nambu-Goldstone boson. 
In some radiative models of neutrino masses, there exists a
Higgs doublet that does not acquire any vacuum expectation value.
This field could be inert and the lightest inert particle could
then be a dark matter candidate. We reviewed these scenarios
in connection with models 
of neutrino masses with right-handed neutrinos and with
triplet Higgs scalars. 

\end{abstract}

\maketitle

\section{Introduction} 

Our understanding about our universe has been improving very fast.
The advancements in technology are contributing to this development
tremendously. Many cosmological models have been ruled out and big-bang
model of cosmology has been accepted as the base model for the evolution
of the universe. The total matter content of the universe is believed to be
same as the critical density of the universe. However, most of this matter
is not visible, about 70 \% of matter is in the form of dark energy or
cosmological constant and another 25 \% of matter is in the form of dark
matter. Only a fraction of the remaining baryonic matter constitute the
entire visible matter. This tells us that the universe is flat and
accelerating at present. 

Although the total amount of visible matter is very small, most of the visible
matter we see around us contains almost no antimatter. This is not easy to 
explain. It is expected that the universe was symmetric with respect to any
conserved charge. So, there would have been equal amount of baryons and
antibaryons and also equal amount of leptons and leptons. Then the observed
baryon asymmetry of the universe requires some explanation. Another puzzle
concerns the dark matter of the universe. Which particles could be the
dark matter candidates. These particles must have been frozen out at some
early time and remained in the background. So, their interaction rate must be
fairly small. On the other hand, if their interaction rate is too small, they
might have overclosed the universe. These dark matter particle must also
have contributed to the formation of the large scale structures in the 
universe, which requires that these particles to be non-relativistic. These
cold dark matter candidates cannot be one of the known particles of the
standard model. A major part of our matter content is in the form of dark
energy or cosmological constant. While this is larger than all other matter
contents, in particle physics standard this is extremely small. The
electroweak phase transition would contribute to the dark energy of the 
universe, which is almost 52 orders of magnitude larger than what is
required at present. Why the observed dark energy is so small requires
a dynamical solution. 

During the past decade many important results came out in neutrino physics.
The atmospheric neutrinos and the solar neutrinos\cite{atm,sol}
 confirmed that the neutrinos
have small masses and there are three generations of neutrinos. 
Including the results from laboratory experimentslike KamLAND \cite{kam}  we now
know the two mass-squared differences and two of three mixing angles fairly
well. The third mixing angle is also constrained by the laboratory experiments.
The mass-squared differences and the mixing angles are:
\begin{eqnarray} 
\Delta m_{atm}^2 &=& 2.1 \times 10^{-3}~{\rm eV}^2 ~~~~~{\rm with} ~~
\sin^2 2\theta_{atm} > 0.92 \nonumber \\[.05in]
\Delta m_{sol}^2 &=& 7.9 \times 10^{-5}~{\rm eV}^2 ~~~~~{\rm with} ~~
\tan^2 \theta_{sol} = 0.4\pm 0.1  \,,
\end{eqnarray}
where $ \theta_{\mu \tau} $ is the mixing angle between $\nu_\mu$ and
$\nu_\tau$ and $ \theta_{e \tau}$ is the mixing angle between $\nu_e$
and one of the other two physical states, which is an admixture of
the states $\nu_\mu$ and $\nu_\tau$. The absolute mass of the neutrinos
have not yet been determined, although there is an upper bound
on the sum over all neutrino masses from cosmology \cite{wm}:
$ \sum_{i=e,\mu,\tau} ~m_{\nu_i} \leq ~ 0.69 ~{\rm eV} .$
The neutrinoless double beta decay also gives
an upper bound on the absolute mass of the neutrinos \cite{ndb},
but this bound is not valid if the neutrinos are Dirac particles.

In this review we shall point out that it is possible to provide
explanation to all the three major issues in cosmology, matter-antimatter 
asymmetry, dark matter and dark energy, in terms of new physics 
coming from models of neutrino masses. While the most natural solution to the
matter-antimatter asymmetry is related to the neutrino masses and called
the leptogenesis, the dark matter and dark energy problems also could have
an explanation coming from neutrino masses. 

\section{Leptogenesis}

There are several mechanisms to generate a baryon asymmetry of the universe
starting from a neutral universe, but most of these mechanisms are
based on one single principle proposed by Sakharov \cite{sakh}. 
It requires three ingredients: {\it i}) Baryon number ($B$) violation,
{\it ii}) $CP$ violation, and
{\it iii}) departure of the $B$-violating interactions from equilibrium.
Instead of $B$ violation, if the interactions violate some combinations
of baryon ($B$) and lepton ($L$) numbers, an asymmetry in that quantum
number will be generated. In grand unified theories all the ingredients
were present \cite{gutbar} and this mechanism could be successfully implemented.
But the asymmetry generated was a $B+L$ asymmetry and it was pointed out
that this asymmetry is washed out by a new process. In the standard model, both
$B$ and $L$ are global symmetries, but the combination $(B+L)$ is broken by
quantum effects arising from anomalous triangle loop diagrams \cite{hooft}.
These anomaly induced $B+L$ violating processes are suppressed
by the quantum tunnelling probability. But at finite temperature,
during the period $10^2 < T < 10^{12}$ GeV,
these interactions become strong in the presence of
some static topological field configuration
called the sphalerons \cite{krs}. This will wash out the $B+L$ asymmetry generated
during the GUT phase transition and make the universe baryon symmetric.
A new proposal \cite{fy} was then made to make use of the sphaleron
transitions to generate a baryon asymmetry from a lepton asymmetry of the 
universe and relate it with the neutrino masses. 

The smallness of the neutrino masses requires lepton number violation
at some large scale. If this lepton number violation generates a lepton
asymmetry of the universe, that will get converted to the required baryon
asymmetry of the universe, before the electroweak phase transition
\cite{fy,ht}. Consider the most
general dimension-5 effective lepton-number violating operator
in the standard model \cite{op}
\begin{equation}
{\cal L}_{Maj} =  \Lambda^{-1} ( \nu \phi^\circ - e \phi^+ )^2 ,
\label{nuop}
\end{equation}
where $ \Lambda $ is some lepton-number violating heavy scale 
and $\phi$ is the usual Higgs doublet scalar. The vacuum expectation
value of $\phi$ will induce a Majorana mass to the neutrinos
\begin{equation}
    {\cal L}_{Maj} = m_\nu \nu_{iL}^T ~C^{-1} \nu_{jL} ,
\end{equation}
with $m_\nu = v^2/\Lambda$. We shall now consider the two simplest 
realizations of this effective operator, which are the see-saw
mechanism \cite{seesaw} and the triplet Higgs mechanism \cite{tripold,ma,trip}.
There are several other interesting models of neutrino masses \cite{oth1}.
In the left-right symmetric extensions of the standard model \cite{lr}
both these contributions are natural for a particular choice of Higgs
scalar \cite{lrnu}. In this article we shall restrict ourselves to only
the seesaw and the triplet Higgs mechanisms. 

To implement the see-saw mechanism \cite{seesaw}, we introduce three right-handed neutrinos 
$N_{Ri},i=e,\mu,\tau$, which are singlets under the standard model. 
The Lagrangian containing the Yukawa interactions of the right-handed neutrinos
is given by,
\begin{equation}
{\cal L}_{int} = h_{\alpha i}~  \overline{\ell_{L \alpha}} \phi ~N_{Ri}
 + M_{ij} ~\overline{(N_{Ri})^c}~ N_{Rj}
\end{equation}
where, $\ell_{L \alpha}$  are the
light leptons,  $h_{\alpha i}$ are the complex  Yukawa  couplings
and $\alpha$ and $i$ are the generation  indices.  
The Majorana masses of the right-handed neutrinos $M_{ij}$ gives
the lepton number violating scale in this case and determines the
mass of the light neutrinos. The Majorana mass of these right-handed
neutrinos allow them to decay into a lepton and an antilepton,
\begin{eqnarray}
  N_{Ri}  &\to& \ell_{jL} + \bar{\phi}, \nonumber \\
   &\to&  {\ell_{jL}}^c + {\phi} .\label{N}
\end{eqnarray}
This lepton number violating decays can generate a lepton asymmetry
of the universe, if there are enough CP violation, which may come from 
the interference of these tree-level decays with some one-loop diagrams,
and these decays satisfy the out-of-equilibrium condition. This lepton
asymmetry will be the same as the $B-L$ asymmetry of the universe,
which is not washed out due to the sphaleron induced processes.
In fact, the sphaleron induced processes will convert this lepton
asymmetry into a baryon asymmetry of the universe. 

There are two sources of CP violation in this scenario :
\begin{itemize}
\item[$(i)$] vertex type diagrams \cite{fy,vertex} which interferes with the tree level
diagram given by figure 1(a). This is similar to the $CP$ violation
coming from the penguin diagram in $K-$decays. For baryogenesis this
diagram was considered in all earlier models. 

\item[$(ii)$] self energy diagrams \cite{self,res} could interfere with the tree level
diagrams to produce CP violation as shown in figure 1(b). This is 
similar to the $CP$ violation in $K-\bar{K}$ oscillation, entering
in the mass matrix of the heavy Majorana neutrinos. In this case the
amount of $CP$ asymmetry could be very large due to resonant enhancement 
\cite{res}, if the masses of the right-handed neutrinos are almost degenerate.

\end{itemize}

\begin{figure}
  \includegraphics[height=.20\textheight]{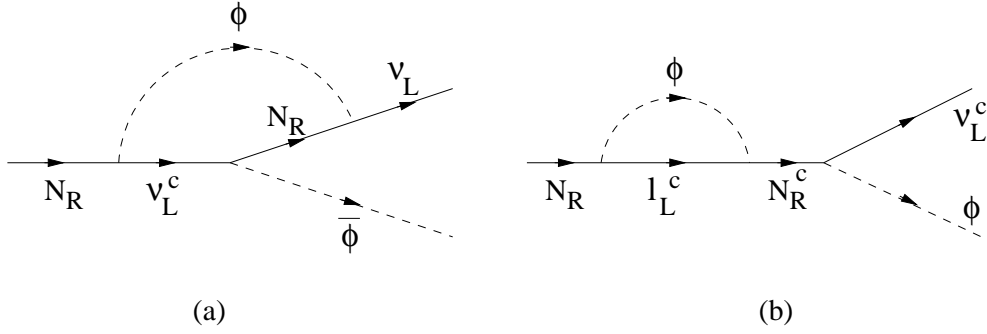}
  \caption{One loop (a) vertex and (b) self energy diagrams, which
interferes with the tree level right-handed neutrino decays to
produce CP violation.
\label{vert}}
\end{figure}

When the decays of the right-handed neutrinos 
satisfies the out-of-equilibrium condition:
\begin{equation}
\Gamma (N \to \ell \bar \phi) < 1.7 \sqrt{g_*} {T^2 \over M_P}
\hskip .5in {\rm at}~T = M_N,
\end{equation}
a lepton asymmetry is generated. Here the right-hand-side 
correspond to the expansion rate of the
universe and $M_P$ is the Planck scale. 
In the case of hierarchical right-handed neutrinos, the amount of lepton asymmetry is
given by 
\begin{eqnarray}
\delta  &=& { \Gamma ( N \to \ell \phi^\dagger) -
\Gamma ( N \to \ell^c \phi) \over \Gamma ( N \to \ell \phi^\dagger) +
\Gamma ( N \to \ell^c \phi)}  \nonumber \\
&=& - {1 \over 8 \pi} \frac{M_1 }{M_2 }
\frac{{\rm Im} [ \sum_\alpha (h_{\alpha 1}^\ast h_{\alpha 2})
 \sum_\beta (h_{\beta 1}^\ast h_{\beta 2}) ] }{ \sum_\alpha 
|h_{\alpha 1}|^2} .
\label{lasy}
\end{eqnarray}
In this expression it has been assumed that the main contribution
to the asymmetry comes from the lightest right handed neutrino 
($N_1$) decay, when the other heavy neutrinos have already decayed away.
The lepton asymmetry thus
generated is same as the $(B-L)$ asymmetry of the universe, since there
is no primordial baryon asymmetry at this time. The sphaleron
interactions now convert this ${(B-L)}$ asymmetry to a baryon asymmetry of the
universe.

We shall now discuss the triplet Higgs model \cite{tripold,ma,trip}. 
In this case one adds two complex 
$SU(2)_L$ triplet higgs scalars ($\xi_a \equiv (1,3,-1); a = 1,2$). 
The $vev$s of the triplet higgses can give small
Majorana masses to the neutrinos through the interaction
\begin{equation}
f_{ij} [\xi^0 \nu_i \nu_j + \xi^+ (\nu_i l_j + l_i \nu_j)/\sqrt 2 
+ \xi^{++} l_i l_j] + h.c.
\end{equation}
If the triplet higgs acquires a $vev$ and break lepton number spontaneously \cite{tripold}, 
then there will be Majorons, which is ruled out by the precision
Z--decay width measurement at LEP. However, if lepton number is violated explicitly
through a trilinear coupling of the triplet Higgs with the standard model Higgs doublet
\begin{eqnarray}
V &=&  \mu (\bar \xi^0 \phi^0 \phi^0 + \sqrt 2 \xi^- \phi^+ \phi^0 + \xi^{--} 
\phi^+ \phi^+) + h.c. \,,
\end{eqnarray}
then the model is not ruled out \cite{ma,trip}. In this case a small $vev$
of the triplet Higgs scalar is ensured naturally. 
If $\langle \phi^0 \rangle = v$ and $\langle \xi^0 \rangle = u$, then the 
conditions for the minimum of the potential relates the $vev$ of the 
two scalars by
$$ u \simeq {{-\mu v^2} \over M^2}, \label{min}\,, $$
where $M$ is the mass of the triplet higgs scalar and 
the neutrino mass matrix becomes $-2 f_{ij} \mu v^2 / M^2 = 2 f_{ij} u$.

\begin{figure}
  \includegraphics[height=.28\textheight]{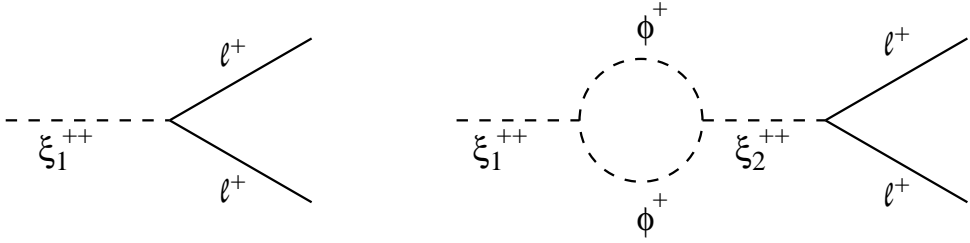}
  \caption{The decay of $\xi_1^{++} \to l^+ l^+$ at tree level and in
one-loop order, whose interference gives ${CP}$ violation.
\label{trl}}
\end{figure}

The lepton number is violated at a large scale $\mu \sim M \gg v$ through
the decays of the triplet Higgs scalars
\begin{equation}
\xi_a^{++} \rightarrow \left\{ \begin{array} {l@{\quad}l} l_i^+ l_j^+ &
(L = -2) \\ \phi^+ \phi^+ & (L = 0) \end{array} \right. .
\end{equation}
$CP$ violation comes from the interference of the tree-level decays
and the self energy diagrams of figure \ref{trl}.
The rate of
$\xi_b \to \xi_a$ no longer remains to be the same as $\xi_b^* \to \xi_a^*$.
Since by $CPT$ theorem $\xi_b^* \to \xi_a^* \equiv \xi_a \to \xi_b$,
it means
\begin{equation}
\Gamma[\xi_a \to \xi_b] \neq \Gamma[\xi_b \to \xi_a] .
\end{equation}
This is a different kind of CP violation compared to the CP violation
in models with right-handed neutrinos. The generated 
lepton asymmetry is given by,
\begin{eqnarray}
\delta &=& { \Gamma ( \xi \to \ell \ell) -
\Gamma ( \xi^\dagger \to \ell^c \ell^c) \over   \Gamma ( \xi \to \ell \ell) +
\Gamma ( \xi^\dagger \to \ell^c \ell^c) } \nonumber \\
&=&
{{Im \left[ \mu_1 \mu_2^* \sum_{k,l} f_{1kl} f_{2kl}^* \right]} \over 
{8 \pi^2 (M_1^2 - M_2^2)}} \left[ {{ M_1} \over 
\Gamma_1}  \right]. \,.
\end{eqnarray}
Depending on the details of the model the scale of leptogenesis is
determined by satisfying the out-of-equilibrium condition. 
The lepton asymmetry thus generated after the Higgs triplets 
decayed away would be the same as the $(B-L)$ asymemtry and the sphaleron
transitions can then convert this asymmetry into a baryon asymmetry
of the universe before the electroweak phase transition, given by
$$ {n_B \over s} \sim {\delta \over 3 g_* K ({\rm ln} K)^{0.6}}. $$
Here $K$ is the Boltzmann suppression factor. 
There are several possible models where the required baryon asymmetry
of the universe could be achieved.

\section{Dark Energy}

Although most of the matter density in the universe is in the form
of dark energy or cosmological constant, the mass scale corresponding
to the present value of the cosmological constant is only
comparable to the neutrino mass scale. On the other hand, the
electroweak symmetry breaking can contribute about 52 orders of
magnitude larger value for the dark energy. It is possible to 
fine-tune some parameters to cancel the large contribution
of the electroweak symmetry breaking, but such fine-tinung has
to be performed to all orders of perturbation theory, which
is unnatural. 
Many attempts to solve the cosmological constant problem has
been made, but the most popular one is 
the quintessence solution \cite{cc07a}. 
There are different quintessence models, which differ in
their predictions for the equation of state of the dark
energy, $\omega = p/\rho$, where $p$ and $\rho$ are the
pressure and the density of the dark energy. But the
present value \cite{wm} of $ \omega = -0.98 \pm 0.12, $
supports the simplest model with $\omega = -1$. 

The quintessence solutions \cite{cc07a}
of the cosmological constant problem
assumes that the mass density of a scalar field $\phi$
gives the dark energy. 
If this field has certain typical
potential that allows slow variation of the field, the dark
energy decreases with time. The scalar field could
be the inflaton field that drives the inflation or
the cosmon field associated with the
the dilatation symmetry \cite{cc07a},
or pseudo Nambu-Goldstone bosons that is presently relaxing
in its vacuum state \cite{cc07b} or
some other scalar field depending on the model.

In recent times the coincidence between the neutrino mass scales
and the scale for cosmological constant has been exploited to
propose a neutrino mass varying solution to the dark energy problem
\cite{mavans}. The 
neutrino mass $m_\nu$ is assumed to be a function of a scalar field,
the acceleron ${\cal A}$, which drives the universe to its present
accelerating phase. In the non-relativistic limit,
the thermal background of the neutrinos and antineutrinos with
a number density $n_\nu$
gives a contribution to the potential amounting to $m_\nu({\cal A}) n_\nu$.
There is also the vacuum energy $V_0(m_\nu)$
attributed to the acceleron ${\cal A}$, so that the effective potential reads
\begin{equation}
    V(m_\mu ) = m_\nu n_\nu + V_0 (m_\nu)\,.
\end{equation}
The relationship of $\omega (t) $ with the scalar potential
is given by
\begin{equation}
\omega +1 = - {m_\nu~ V_0^\prime (m_\nu) \over V} =
{\Omega_\nu \over \Omega_\nu + \Omega_{DE}} .
\end{equation}
The observed value of $\omega \approx -1$ implies that the energy
density in neutrinos is much smaller than the total energy
density, which corresponds to $m_\nu ~V_0^\prime \ll
V(m_\nu)$ and hence a flat potential, which could be
taken to be
\begin{equation}
V_0(m_\nu) = \Lambda^4 \log \left(1 + \left|{M_1({\cal A}) \over \mu}
\right| \right) .
\end{equation}
The form of $M_1 ({\cal A})$ determines the dynamics of the model. 

We shall now discuss a possible source of the acceleron field,
which could be a pseudo-Nambu-Goldstone
boson (pNGB) originating from lepton flavor violation \cite{pngb}.
Consider a a two-generation example with two right-handed neutrinos
$N_i, i=1,2$, whose Yukawa couplings 
with some scalar singlet fields $\Phi_i, i=1,2$ is given by
\begin{equation}
\L_N = { 1 \over 2} \alpha_1 \bar N_1 N_1^c \Phi_1 +
{ 1 \over 2} \alpha_2 \bar N_2 N_2^c \Phi_2 .
\end{equation}
The transformation of these fields under 
a global $U(1)_1 \times U(1)_2$ symmetry of the model can be read as
$N_1 \equiv (1,0)$, $N_2 \equiv (0,1)$, $\Phi_1 \equiv (2,0)$ and
$\Phi_2 \equiv (0,2)$ respectively. The vacuum expectation values
of the fields $\Phi_i$ break both these symmetries leading to two
Nambu-Goldstone bosons. One of these Nambu-Goldstone bosons remain
massless, which corresponds to total lepton number, which is
the singlet Majoron. The other Nambu-Goldstone boson eventually
picks up a finite and small mass and become pseudo-Nambu-Goldstone boson
(pNGB). 

We further assume that the left-handed doublets
$\ell_{\alpha L}^T \equiv
\pmatrix{\nu_\alpha & \alpha}, \alpha = e, \mu$ transform
under $U(1)_1 \times U(1)_2$ as (1,0) and (0,1), respectively.
If the standard model Higgs doublet is allowed to have the usual 
Dirac Yukawa couplings with the neutrinos, then that will break the
global symmetry explicitly. However, this will spoil the renormalizability
of the theory and the pNGB will receive infinite corrections to its mass
and will have a mass of the order of $\langle \Phi \rangle$. So, all
Yukawa couplings should remain invariant under the global
$U(1)_1 \times U(1)_2$ symmetry, which could be achieved by
introducing three Higgs doublet fields $H_a, a = 0,1,2$,
which transforms under $U(1)_1 \times U(1)_2$ symmetry as
$(0,0), (1,-1)$ and $(-1,1)$, respectively. The Yukawa
couplings 
\begin{eqnarray}
{\cal L}_{mass} &=& f_{11} \bar N_1 \ell_1 H_0 + f_{12} \bar N_1
\ell_2 H_1
+f_{21} \bar N_2 \ell_1 H_2 + f_{22} \bar N_2 \ell_2 H_0 
\end{eqnarray}
will then give the Dirac masses for the neutrinos
$$ \L_{mass} = m_{ij} \bar N_i \nu_j, $$
which breaks the $U(1)_1 \times U(1)_2$ symmetry softly. 
This will then give a finite small mass to the pNGB, which
can be evaluated from the one-loop Coleman-Weinberg potential. 
This light pNGB with mass of the order of the neutrino masses
$m_{\cal A} \sim m^2/M$ could then become the acceleron field. 

After proper transformations and taking care of the redundant phases,
we can write down the neutrino masses as
\begin{eqnarray}
{\cal L}_\mu &=& {1 \over 2} M_i({\cal A}) \bar{N_i^c} N_i
+ m_{ij} \bar N_i \nu_j + H.c.
\label{full1}
\end{eqnarray}
The right-handed neutrino mass shows explicit dependence on the
acceleron field and $M_i({\cal A})$ is specified. 
Then the effective neutrino mass also varies $m_\nu ({\cal A}) = m^T M^{-1} ({\cal A})
m$, and we finally arrive at 
\begin{equation}
- {\cal L}_{eff} = m_{ij}^T [M_i ({\cal A})]^{-1} m_{ij}
~ \nu_i \nu_j +H.c. + \Lambda^4 \log (1 + |{M_1({\cal A}) / \mu} | ) ,
\end{equation}
whose minima corresponds to $\omega = -1$.

The models of mass varying neutrinos have some more problems \cite{peccei},
which are being taken care of in some variants of the model
\cite{modelsown,models}. The naturalness problem could be softened
drastically in models with triplet Higgs scalars
\cite{modelsown}, which provides interesting phenomenology at LHC.
Like any models of quintessence, the models of mass varying neutrinos
are not yet fully consistent. However, the rich phenomenological
implication of this model makes it interesting. One of the implications
is the existence of a long range force
that could be tested in the near future.

\section{Dark Matter}

There are many dark matter candidates, including the lightest
supersymmetric particle, which is usually a neutralino. Here
we shall restrict ourselves to the dark matter candidate that
appears from models of neutrino masses \cite{dm}. One natural explanation
of the smallness of neutrino masses is due to radiative models. 
In one such model one introduces an additional Higgs doublet $\eta$. 
An extra $Z_2$ symmetry prevents this Higgs from coupling with
the quarks. The right-handed neutrinos ($N_i$, i = 1,2,3) are
also odd under the $Z_2$ symmetry so that the 
scalar potential for the fields $\phi$ and $\eta$
and the Yukawa interactions of the right-handed neutrinos are
given by
\begin{eqnarray}
V &=& \mu_1^2 |\phi|^2 + \mu_2^2 |\eta|^2 + \lambda_1 |\phi|^4
+ \lambda_2 |\eta|^4 + \lambda_3 |\phi \eta|^2
+ \lambda_4 |\phi^\dagger \eta|^2 + {\lambda_5 \over 2}
[(\phi^\dagger \eta)^2 + H.c.]. \nonumber \\[.15in]
{\cal L}_{Y} &=& h_{\alpha i} ~( \nu_\alpha \eta^\circ - \ell_\alpha
\eta^+ )N_i + H.c. + M_i N_i N_i \\[-.15in]
\end{eqnarray}
We now assume that only the standard model Higgs $\phi$ acquires
a $vev$, $\langle \phi \rangle = v$ and
$ \langle \eta \rangle = 0$. 
The fields $\eta^+$ and $\eta^\circ_I$ will become the longitudinal
modes of the gauge bosons making them massive and the usual
physical Higgs scalar $\phi^\circ_R$ can now become as heavy as
400-600~GeV. The scalars $\eta^+$, $\eta^\circ_R$ and $\eta^\circ_I$
do not interact with the light fermions and remain inert. It is now
possible to choose the coupling constants of the scalar potential,
which will make one of the neutral components of $\eta$ to be the
lightest inert particle (LIP). The $Z_2$ symmetry will make the
LIP to be a stable particle and hence a candidate for the dark
matter of the universe. 
The parameters of the scalar potential can be constrained by the
present limits on the dark matter. The strongest requirement
is $\lambda_5 \neq 0$, which is also required for the neutrino masses. 
The neutrino mass matrix is then given by
\begin{equation}
{\cal M}_{\nu \alpha \beta} = \sum_i {h_{\alpha i} h_{\beta i} M_i
\over 16 \pi^2} \left[ {m_R^2 \over m_R^2 - M_i^2} {\ln} {m_R^2 \over
M_i^2} - {m_I^2 \over m_I^2 - M_i^2} {\ln} {m_I^2 \over
M_i^2} \right] \,.
\end{equation}
We assumed the Majorana masses of the right-handed neutrinos are real
and diagonal. The lightest inert particle can now have a mass in the
range of 50-100 GeV and can be a candidate for the cold dark matter
in the universe. 

\section{Summary }

Our present understanding of our universe suggests that about 70\%
of matter is in the form of dark energy, about 25\% is non-baryonic
nonrelativistic dark matter
and a fraction of the remaining matter is visible matter and there
is very little scope of antimatter. Considering that the total
matter is about the critical matter density of the universe, this
points towards the neutrino mass scale for all the matter. There 
are now suggestions to explain why there are more matter compared
to antimatter, what are the dark matter and how to explain why the
dark energy is so small, all in terms of some properties of neutrino
physics. We discussed these solutions in a couple of models of 
neutrino masses, namely the one with right-handed neutrinos and
the triplet Higgs scalars.

\newcommand{\ib}{{\it ibid.}}
\newcommand{\np}[1]{{ Nucl. Phys. }{\bf #1}}
\newcommand{\pl}[1]{{ Phys. Lett. }{\bf #1}}
\newcommand{\pr}[1]{{ Phys. Rev. }{\bf #1}}
\newcommand{\prl}[1]{{ Phys. Rev. Lett. }{\bf #1}}
\newcommand{\zp}[1]{{ Z. Phys. }{\bf #1}}
\newcommand{\prep}[1]{{ Phys. Rep. }{\bf #1}}
\newcommand{\rmp}[1]{{ Rev. Mod. Phys. }{\bf #1}}
\newcommand{\ijmp}[1]{{ Int. J. Mod. Phys. }{\bf #1}}
\newcommand{\mpl}[1]{{ Mod. Phys. Lett. }{\bf #1}}
\newcommand{\ptp}[1]{{ Prog. Theor. Phys. }{\bf #1}}
\newcommand{\arnps}[1]{{ Ann. Rev. Nucl. Part. Sci. }{\bf #1}}
\newcommand{\PL}[1]{{ Phys. Lett. }{\bf #1}}
\newcommand{\PR}[1]{{ Phys. Rev. }{\bf #1}}
\newcommand{\PRL}[1]{{ Phys. Rev. Lett. }{\bf #1}}
\newcommand{\ZP}[1]{{ Z. Phys. }{\bf #1}}
\newcommand{\NP}[1]{{ Nucl. Phys. }{\bf #1}}

\end{document}